%% file: main.tex
\documentclass[sigconf]{acmart}
\usepackage{graphicx}

\usepackage{ifthen}
\newboolean{techreport}
\usepackage{caption}
\usepackage{multirow}
\usepackage{subcaption}
\usepackage{hyperref}
\usepackage{amsmath}
\usepackage{tikz}
\usepackage[]{footmisc}
\usepackage{graphicx}
\usepackage{pifont}

\usepackage[ruled,linesnumbered,vlined]{algorithm2e}
\usepackage{amsmath,amstext}
\let\oldnl\nl
\newcommand{\nonl}{\renewcommand{\nl}{\let\nl\oldnl}}
\usepackage{tikz}

\definecolor{codegreen}{rgb}{0,0.6,0}
\definecolor{codegray}{rgb}{0.5,0.5,0.5}
\definecolor{codepurple}{HTML}{C42043}
\definecolor{backcolour}{rgb}{1,1,1}

\newcommand\numberstyle[1]{%
    \footnotesize
    \color{codegray}%
    \ttfamily
    \ifnum#1<10 0\fi#1 |%
}

\usepackage{enumitem}

\renewcommand\footnotetextcopyrightpermission[1]{} 

\begin{document}
\setboolean{techreport}{true}

\title{Dronevision:  An Experimental 3D Testbed for Flying Light Specks}
\author{Hamed Alimohammadzadeh, Rohit Bernard, Yang Chen, Trung Phan, Prashant Singh, Shuqin Zhu, Heather Culbertson, Shahram Ghandeharizadeh}
\affiliation{%
  \institution{University of Southern California}
    \city{Los Angeles}
  \country{USA}
}


\begin{abstract}
\input abs
\end{abstract}

\maketitle
\pagestyle{plain} 

\section{Introduction}\label{sec:intro}
\input{intro}

\section{Flying Light Specks}\label{sec:fls}

\input{fls}

\section{User Interaction}\label{sec:interaction}
\input{haptics}

\section{Multimedia Systems Challenges}\label{sec:mmchallenges}
A DV raises many interesting multimedia systems challenges.
The characteristics of a challenge may vary for different applications.
This section presents two challenges. 

\subsection{FLS Localization}\label{sec:localize}
\input{localize}

\subsection{3D Acoustics}\label{sec:audio}
\input{audio}

\section{Related Work}\label{sec:related}
\input{related}

\section{Conclusions and Current Efforts}\label{sec:conc}
\input{dv}

\section{Acknowledgments}
This research was supported in part by the NSF grant IIS-2232382.

\bibliographystyle{ACM-Reference-Format}
\bibliography{refs}  

\end{document}

%% file: abs.tex
Today's robotic laboratories for drones are housed in a large room.  At times, they are the size of a warehouse.  These spaces are typically equipped with permanent devices to localize the drones, e.g., Vicon Infrared cameras.  Significant time is invested to fine-tune the localization apparatus to compute and control the position of the drones.  One may use these laboratories to develop a 3D multimedia system with miniature sized drones configured with light sources.  As an alternative, this brave new idea paper envisions shrinking these room-sized laboratories to the size of a cube or cuboid that sits on a desk and costs less than 10K dollars.  The resulting Dronevision (DV) will be the size of a 1990s Television.  In addition to light sources, its Flying Light Specks (FLSs) will be network-enabled drones with storage and processing capability to implement decentralized algorithms.  The DV will include a localization technique to expedite development of 3D displays.  It will act as a haptic interface for a user to interact with and manipulate the 3D virtual illuminations.  It will empower an experimenter to design, implement, test, debug, and maintain software and hardware that realize novel algorithms in the comfort of their office without having to reserve a laboratory.  In addition to enhancing productivity, it will improve safety of the experimenter by minimizing the likelihood of accidents.  This paper introduces the concept of a DV, the research agenda one may pursue using this device, and our plans to realize one. 

%% file: intro.tex

\begin{figure}
\centering
\includegraphics[width=\columnwidth]{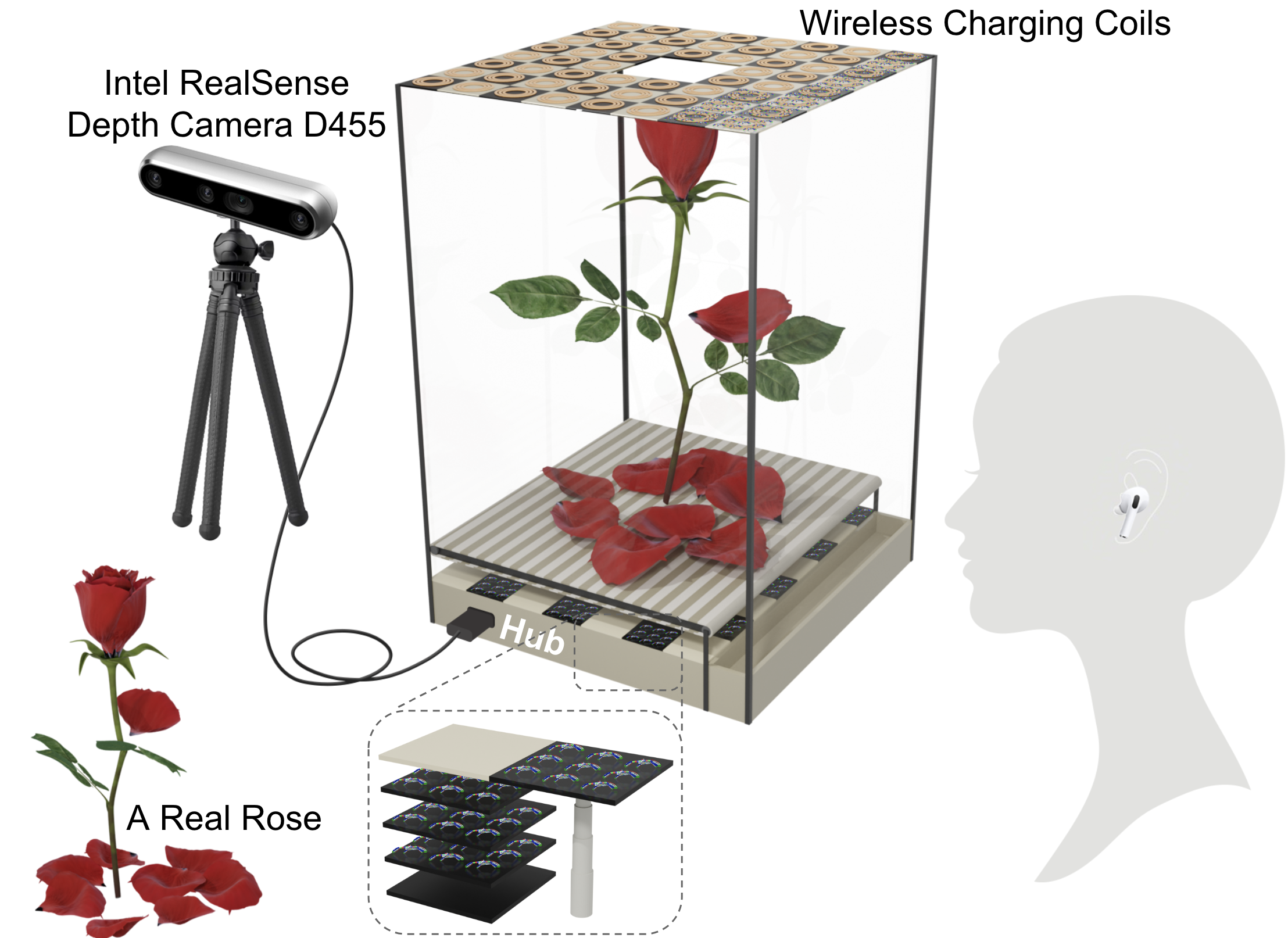}\hfill
\caption{A DV with glass panes.}
\label{fig:depthcam}
\end{figure}

\begin{figure}
\centering
\includegraphics[width=0.9\columnwidth]{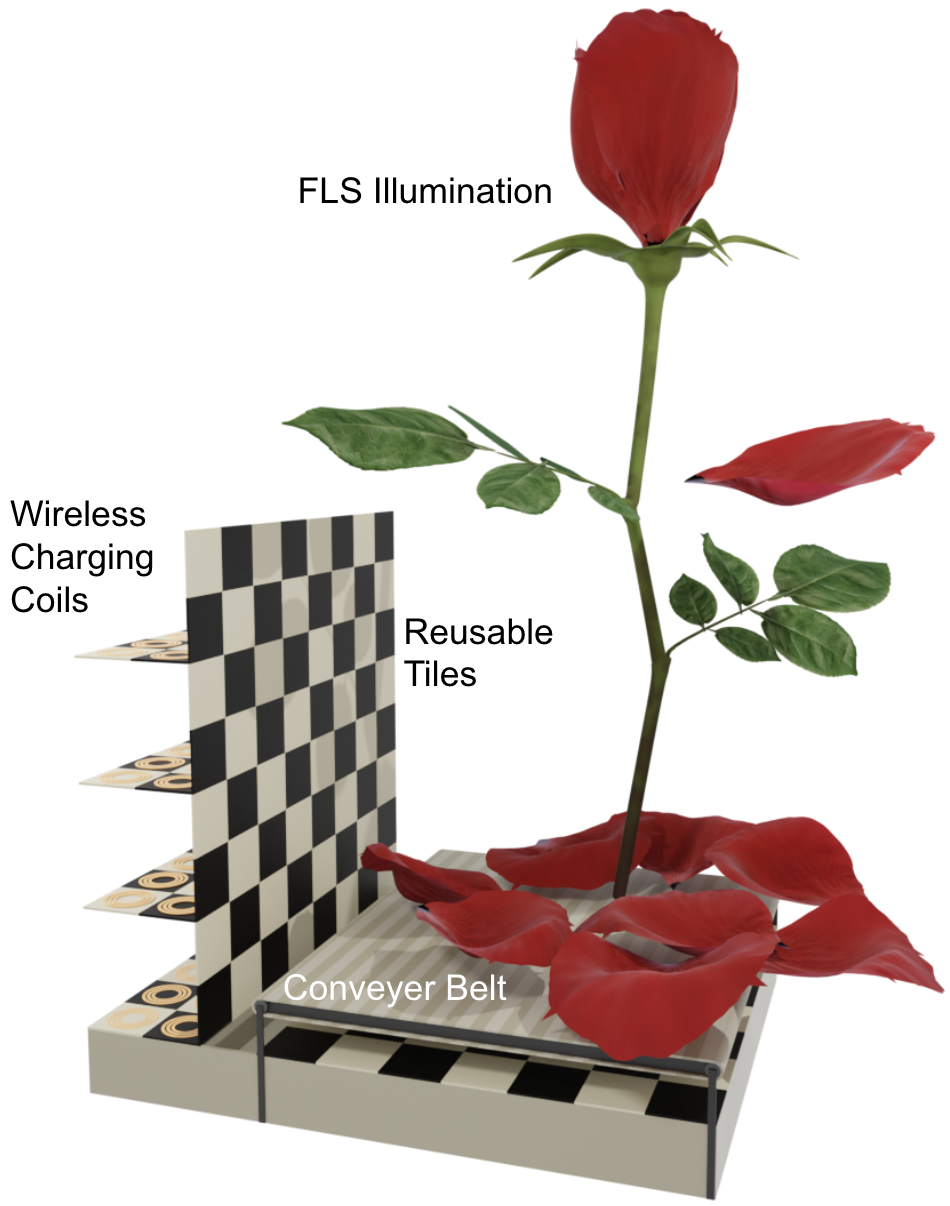}\hfill
\caption{A DV rendering a large illumination.}
\label{fig:wall_hanging_display}
\end{figure}

A Dronevision, DV, is a 3D display to design and implement next generation multimedia applications using Flying Light Specks, FLSs.
An FLS is a miniature sized drone with one or more light sources with adjustable brightness~\cite{shahram2021,shahram2022,shahram2022b}.
A swarm of cooperating FLSs will render 3D shapes, point clouds, and animated sequences in a fixed volume, a 3D display.
This {\em Brave New Idea} (BNI) envisions shrinking today's room sized robotics laboratories to the size of a cuboid that fits on a user's desk for multimedia content creation and rendering. 
Figure~\ref{fig:depthcam} shows a DV setup configured for visual display only.
Four glass panes enhance the user's safety by preventing rogue FLSs from coming to contact with a user and injuring them.
Once the rendering of an illumination is verified to be safe,
the user may remove the glass panes.
The DV may illuminate the rose beyond the size of its base, see Figure~\ref{fig:wall_hanging_display}.
Moreover, a user may touch and interact with the illumination.

For 3D media content, a DV is a low cost alternative to an expensive robotics laboratory shared by different research groups.
It will provide an experimenter with a dedicated 3D testbed to debug and evaluate their hardware and software solutions.
It will enable a content creator to design and implement diverse multimedia applications that range from healthcare to entertainment.
With healthcare, a designer may develop a 3D FLS display to illuminate 3D MRI scans of a patient and their organs, empowering a physician to separate the different organs (illuminations) and examine them in real time.  
With entertainment, the designer may illuminate characters of multi-player games such as Minecraft and Fortnite to come alive and interact with one another on a table top.
In addition to improving productivity, a DV enhances the safety of both the experimenter and the content creator.


Figure~\ref{fig:depthcam} shows a DV displaying an illuminated rose with a falling petal; its original may be captured using an Intel RealSense depth camera\footnote{The use of a camera is for illustration purposes only.  The shown rose is a mesh file consisting of 65K points created using Blender~\cite{shahram2022,mmsys2023}.}.
The top panel of the DV consists of wireless charging stations.
FLSs with a low remaining flight time fly through the square hole and land on a charging coil to charge their battery.
Hundreds of FLSs may land on a charging coil and simulataneously charge their battery.

The floor of the DV consists of a Hub, Tiles, Terminus, and a conveyor belt.  
Each rod on the side acts as an elevator that delivers fully-charged FLSs to the base of the DV, where they are stored on the black tiles for future deployment.

The Hub is comparable to today's servers and provides both wired and wireless connectivity to peripheral devices such as a depth camera and earbuds.
Black and white tiles are modular pieces that can be plugged into other tiles to construct DVs of different lengths and depths.
Black tiles serve both as a hangar and a dispatcher of FLSs, which are moved up and down using an elevator, and white tiles are rigid pieces that keep the floor together.
Black tiles containing FLSs may be stacked below the white tiles, see Figure~\ref{fig:depthcam}. When an empty black tile is pushed below a white tile, a black tile with FLSs is   
pushed onto the elevator to bring its FLSs to the floor of the DV; these FLSs are then dispatched to illuminate virtual objects.

Failed FLSs will fall onto a conveyor belt, which is a garbage collector that deposits failed FLSs to a Terminus.
The Terminus is an empty row on the DV floor, e.g., the row closest to the user (the back panel) in Figure~\ref{fig:depthcam} (\ref{fig:wall_hanging_display}).
The conveyor belt protects FLSs sitting on black tiles from falling FLSs that have failed. 

The Hub stops the conveyor belt before deploying one or more FLSs atop a black tile. A deployed FLS flies below the conveyor belt and past the Terminus to navigate to its coordinates in the DV's display volume.
Stopping the conveyor belt minimizes the likelihood of a failed FLS hitting and damaging a deployed FLS.
Similarly, by placing the charging stations at the top of the DV, the likelihood of failed FLSs hitting and damaging a functional FLS that is in transit to a charging station is minimized. 

A DV will be both modular and configurable.
Once an experimenter is satisfied with the safety of their setup in Figure~\ref{fig:depthcam}, they may remove the glass panes and place the top as a back panel in front of the Terminus, enabling the DV to render illuminations that extend beyond the length and width of its base and allowing the user to physically interact with the illuminations (Figure~\ref{fig:wall_hanging_display}). 
The haptic feedback provided by the illumination will include a combination of kinesthetic and tactile sensations. The kinesthetic (force) feedback will enable a user to experience the shape and hardness of an illuminated object.
For example, a user may grasp a rose petal by its sides using their fingers.  The user may press their fingers together to bend the rose petal to experience its stiffness.
The tactile (skin-based) feedback will enable a user to experience the object's surface properties.
For example, FLSs with electrotactile actuators may provide the sharp sensation a user would expect when running their finger over a thorn on the virtual rose of Figure~\ref{fig:wall_hanging_display}.


The {\bf contributions} of this paper include:
\begin{itemize}
\item DV as an experimental 3D platform to design and implement next generation multimedia applications using FLSs (Sections~\ref{sec:intro} and~\ref{sec:fls}).

\item Use of DV for haptic interactions (Section~\ref{sec:interaction}).

\item Two multimedia systems research challenges raised by a DV 
(Section~\ref{sec:mmchallenges}).

\end{itemize}
We present related work in Section~\ref{sec:related} and brief conclusions and our current efforts to realize a DV in Section~\ref{sec:conc}.

%% file: fls.tex


An FLS is a miniature sized drone configured with light sources.  Each FLS is too simple to illuminate a point cloud by itself or facilitate complex and noticeable haptic interactions.  The inter-FLS relationship effect of an organizational framework will compensate for the simplicity of each individual FLS, enabling a swarm of cooperating FLSs to illuminate complex shapes and render haptic interactions.
Therefore, we stipulate that an FLS must include communication in order to coordinate and cooperate with other FLSs, storage to memoize its assigned lighting tasks at specific coordinates, processing to implement algorithms, and a wide variety of sensors for localization, collision avoidance, and haptic interactions.

Eleven quadrotors that measure between 1.8 in (45 mm) and 5 in (127 mm) from motor shaft to motor shaft and weight from 0.4 oz (11.5 g) to 2.5 oz (72 g) are evaluated in~\cite{deters2018}.
A survey of 93 rotorcrafts is presented in~\cite{ardelean}, which classifies the rotorcrafts along 12 dimensions including rotor type, application, battery capacity among others.
It uses regression to establish relationships between size and performance parameters for the preliminary design of rotorcraft.
This study found a wide variety of small drones that are already mass-produced as consumer products, meaning that miniaturization of components has already received significant attention. 
This miniaturization, along with advancements in batteries and autopilots, shows the feasibility of smaller and lighter FLSs.

\begin{figure}
\centering
\includegraphics[width=0.8\columnwidth]{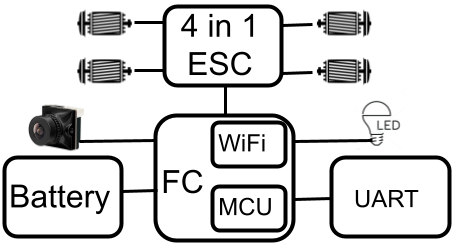}\hfill
\caption{Today's Architecture.}
\label{fig:archdrone}
\end{figure}

Figure~\ref{fig:archdrone} shows the architecture of today's drone electronics~\cite{speedybee2023}.
One of the most important components, the Flight Controller (FC), stabilizes the FLS 
to perform precise flight maneuvers. 
Today’s FCs are circuit boards that adjust the speed of the motors to move the drone in a desired direction.
They interface with an Electronic Speed Controller, ESC, that connects 
to the motors, controlling the speed at which the motors must rotate for the applied throttle.
A 4-in-1 ESC allows an FC to control all four motors of a quadrotor. 

An FC is equipped with sensors that detect the drone’s movements. 
Basic FC sensors include gyroscopes (gyro) and accelerometers (acc).
While the gyro is used to measure angular velocity, the acc measures linear acceleration.
Other sensors may include barometric pressure sensors and compasses (magnetometer).  
Today's FC may serve as a hub for other drone peripherals like LED light sources, cameras, and video transmitters (VTX), see Figure~\ref{fig:archdrone}.
A current trend is smaller FCs with more features.

Different FC boards may require different firmware.  Configuring the FC’s firmware is the process of adjusting settings of its Proportional–Integral–Derivative (PID) controller, RC rates which determine how rapidly the drone rotates about an axis, and others to achieve desired flight characteristics.  It strives to harness the full potential of the FC for a target application.
A firmware setting may use only one sensor such as the gyro\footnote{This is the Acro Mode setting of Betaflight firmware.} or require multiple sensors\footnote{Examples include Betaflight’s Angle Mode, Horizon Mode, and Rescue mode.}, such as both gyro and acc.

An FC may include a wireless networking card, and uses a microcontroller unit, MCU, to store firmware data and perform complex computations.
At the time of this writing, MCU speeds range from 72 MHz to 480 MHz with memory sizes ranging from 128 KB to 2 MB.
The MCUs are evolving rapidly to provide faster speeds and higher memory capacities, retiring older models.

An FC has a Blackbox feature that logs attitude of the drone, gyro sensor measurements, RC commands, motor outputs, etc., which is useful for tuning and troubleshooting an FLS.  It requires the FC to have either a flash memory chip (e.g., 
NOR flash) or a SSD card recorder.  The rate of logging is an adjustable parameter.  Visualization tools for the Blackbox logs enable a user to select the traces and generate graphs of interest.  Example timeline graphs include FC throttle levels, attitude, motor outputs, and PID controller outputs.

Today’s FCs use a hardware serial interface named Universal Asynchronous Receiver/Transmitter (UART) that allows external devices to be connected to the FC.  Examples include serial radio receivers, telemetry, race transponders, and VTX control.  It is possible to create additional UART ports in the firmware, using the MCU to emulate multiple UARTs.  This feature is called SoftSerial.

\begin{figure}
\centering
\includegraphics[width=\columnwidth]{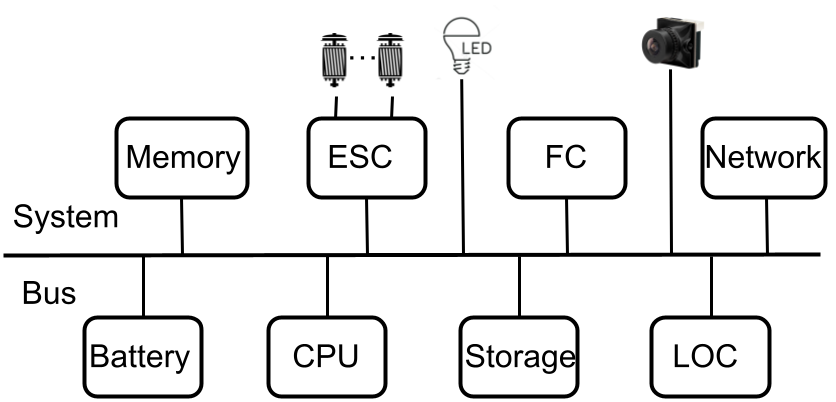}\hfill
\caption{Future FLS Architecture.}
\label{fig:archfls}
\end{figure}

\noindent{\bf Future FLS Architectures:}
We envision the future architecture of an FLS will include a System-on-Chip, SoC, and disaggregated designs.
The SoC will resemble that of Figure~\ref{fig:archdrone} with additional functionality implemented in FC.  
The disaggregated design is shown in Figure~\ref{fig:archfls}.
It envisions a plug-n-play system bus that empowers an experimenter to use with alternative devices.
The tradeoff between the two include weight, size, and modularity with the ability to switch components.
The SoC will almost certainly be lighter and more compact. 
However, it will be difficult (if not impossible) to change one or more of its components.
For example, it may be challenging to replace its WiFi hardware component with the power efficient ZigBee (802.15.4, 2.4 GHz).
This may be ideal for a product offering such as a next generation gaming display.  
However, for research and evaluation purposes, a disaggregated design may be preferable as it empowers an experimenter to evaluate alternative hardware components and their tradeoff when designing and implementing different algorithms.
In essence, one may use the disaggregated design to identify components of a SoC.
Below, we describe the disaggregated architecture.

There are several differences between the architecture of Figure~\ref{fig:archdrone} and the disaggregated architecture of Figure~\ref{fig:archfls}.
First, the disaggregated architecture requires external devices to interface with the system bus directly, making a UART redundant.
Second, it replaces 
MCU with a Central Processing Unit, CPU.
The CPU will host a general purpose operating system.
It will enable an experimenter to implement and evaluate diverse algorithms such as collision detection and avoidance~\cite{clearpath2009,collisionfree2012,collisionfree2015,collisionavoidance2018,ReactiveCollisionAvoidance2008,ReactiveCollisionAvoidance2011,ReactiveCollisionAvoidance20112,reactiveColAvoidance2013,downwash1,downwash3,dcad2019,Engelhardt2016FlatnessbasedCF,navigation2017,reactiveColAvMorgan,reactiveColAvBaca,reactiveColAvMorganJ,speedAdjust2021,gameCollisionAvoidance2020,gameCollisionAvoidance2017,preiss2017,Ferrera2018Decentralized3C,planning2019,preiss2017whitewash,opticalpositioning1,navigate2014}, a continuous battery charging technique such as STAG~\cite{shahram2022}, localization~\cite{chenSurvey2022,gusurvey2009,luca2014,zafari2019}, 3D audio~\cite{shahram2021}, among others~\cite{shahram2021}.  
A Localization device, LOC, will enable the experimenter to plug-in alternative localization sensors such as Bluetooth~\cite{bluetooth1,bluetooth2}, Wi-Fi~\cite{wifi1,wifi2}, RFID~\cite{rfid1,rfid2,rfid3,rfid4}, UWB~\cite{uwb1,uwb2,uwb3}, Lidar~\cite{lidar1,rgblidar}, RGB cameras~\cite{rgb1,rgb2,rgb3,rgblidar} and infrared~\cite{opticalpositioning1,opticalpositioning2,preiss2017}.
They will be used by the software hosted on the local CPU to compute the position of an FLS relative to other FLSs and the display volume.

A challenge of Figure~\ref{fig:archfls} is how to realize a layout that is light, compact, economical, and adjustable by an experimenter with minimal effort.
While the architecture of Figure~\ref{fig:archdrone} is not adjustable, it is lightweight and compact.
For example, SpeedyBee\footnote{F7 identifies a 216 MHz MCU with either 0.5 or 1 MB of memory.} F7~\cite{speedybee2023} stacks an FC with a 4 in 1 ESC that weighs 19.2 grams with dimensions of 45.6mm (Length) x 40mm (Depth) x 8.8mm (Height) at \$119.
Ideally, a disaggregated design should closely approximate these specifications, and its future SoC implementations should be orders of magnitude smaller, lighter, and cheaper. 

Downwash is the rapid displacement of air by an FLS's rotors.
It may impact the position of other FLSs, resulting in unstable visuals and inconsistent haptic feedbacks.
An FLS may use an existing method such as a neighbor-aware states to train its policy to counter downwash~\cite{Shi2020, Riviere_2020}. Moreover, FLSs may allocate extra thrust to counteract disturbances.
This will require an FLS to provision its thrust to implement other tasks such as haptic feedback. 

An FLS has a fixed flight time on a fully charged battery.
An alternative is to use a continuous power source such as a laser power beam~\cite{Achtelik} or a tethered FLS~\cite{Frédéric} with an umbilical link to a power source.
It may also be possible to have a laser beam charge an FLS mid-flight~\cite{Ouyang}. 
A charging software solution such as STAG~\cite{shahram2022} also benefits from a hardware solution that enables an FLS to switch its depleted battery with a fully charged one~\cite{Swieringa}.

%% file: haptics.tex
FLSs will facilitate both kinesthetic (force) and tactile (skin-based) feedback to a user.
While the former provides the shape and hardness of a virtual object, the latter renders an object's surface properties.
Both motivate the disaggregated FLS architecture of Figure~\ref{fig:archfls} with diverse plug-in devices.
To illustrate, textures for different surfaces may be rendered using different vibration actuators~\cite{culbertson2016importance}.
An actuator may be a plug-in device to the FLS system bus, enabling an experimenter to evaluate the actuator and its tactile feedback.  

\begin{figure}
  \begin{center}
    \includegraphics[width=0.9\columnwidth]{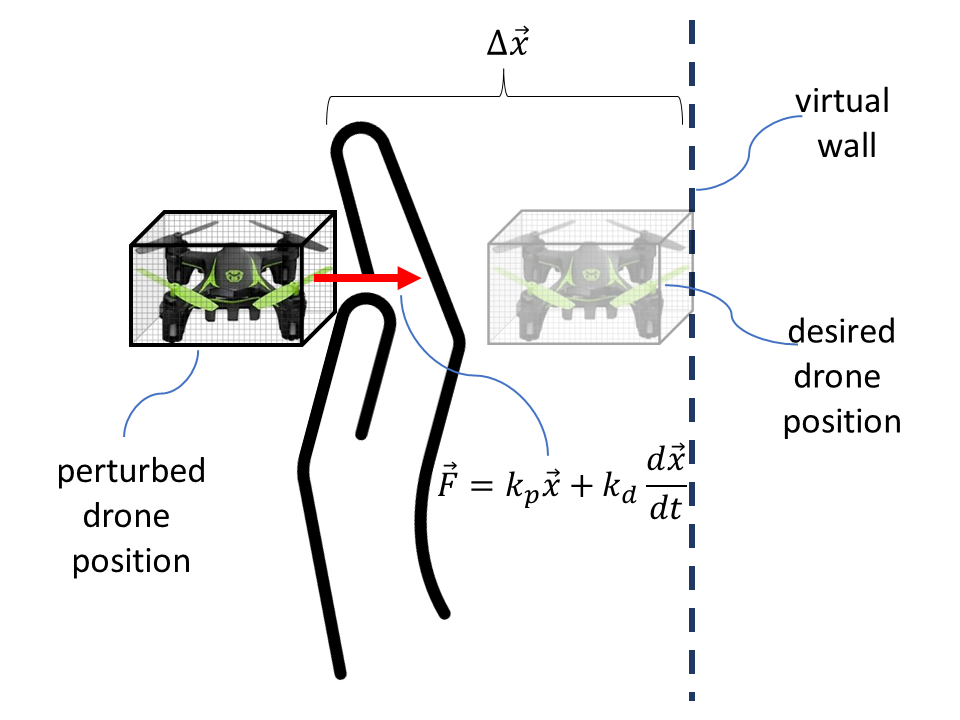}
  \end{center}
  \caption{Rendering scheme for a single FLS. Force is applied to the user proportional to the distance that the FLS is perturbed away from a setpoint location.}\label{fig:force}
\end{figure}

An FLS acts as an encountered-type haptic device.
Traditional kinesthetic 
feedback haptic rendering uses a vector-field approach~\cite{massie1994phantom} where the force to be displayed to the user is calculated based on their position in the 3D environment and geometric models of the virtual objects. Our FLS-based system will take a similar approach and calculate the desired force based on the user's position and the 3D model of the object(s) to be rendered.
For example, in displaying a virtual wall, the user is free to move their hand around the environment until their position penetrates inside the virtual wall, at which time the FLS would make contact with their hand to apply a force and prevent them from penetrating further into the virtual wall (Fig.~\ref{fig:force}). Using measurements of the FLS's position and a 3D model of an object, we can determine how far the user has penetrated into the virtual object to calculate the amount of force to display to the user.

A standard FC uses PID control to keep the drone at the desired location in 3D space. However, we found in our tests that the integral \textit{I} portion of the controller caused instability issues during user interactions because of wind-up, which resulted in the FLS being unable to remain steady when perturbed by the user. Altering the FC to use only PD control removes this source of instability. Therefore, the force displayed to the user during the interaction is proportional to the distance that the user has perturbed the drone away from its setpoint (i.e., the edge of the virtual object) and the speed of the push, $\vec{F}=k_p\vec{x}+k_dd\vec{x}/dt$.

With grounded impedance force-feedback haptic devices, the calculated force is then converted to motor torque using the device's calibrated Jacobian matrix so that the desired force can be displayed to the user.
Since the controllable output of the drones is thrust, displaying a set force to the user requires careful calibration between the commanded thrust and the resulting force. Previous researchers have completed this calibration for single axes of thrust/force for the Parrot AR.Drone~\cite{abdullah2018hapticdrone}. Displaying more complex objects and interactions in 3-dimensions requires a more complete calibration of the system.


\noindent{\bf Force Amplification:}
The size of the FLS plays an important role in its dynamics and the force that can be displayed to the user. The FLS's motors determine the highest RPM that can be generated, and the propellers' ratio of pitch:diameter affects the force per RPM generated from the air pushing downwards. A larger FLS usually means larger diameter propellers, leading to higher forces at a given RPM. However, the force that we will be able to output with an FLS will be significantly less than that of a traditional kinesthetic haptic device; for example, the common 3D Systems Touch device has a maximum force output of 3.3~N. The modality of human sensing also varies based on the level of force provided, with small forces (<1 N) being sensed as tactile information through the skin rather than kinesthetic information through the muscles~\cite{jones2006human}.

\begin{figure}
  \begin{center}
    \includegraphics[width=0.9\columnwidth]{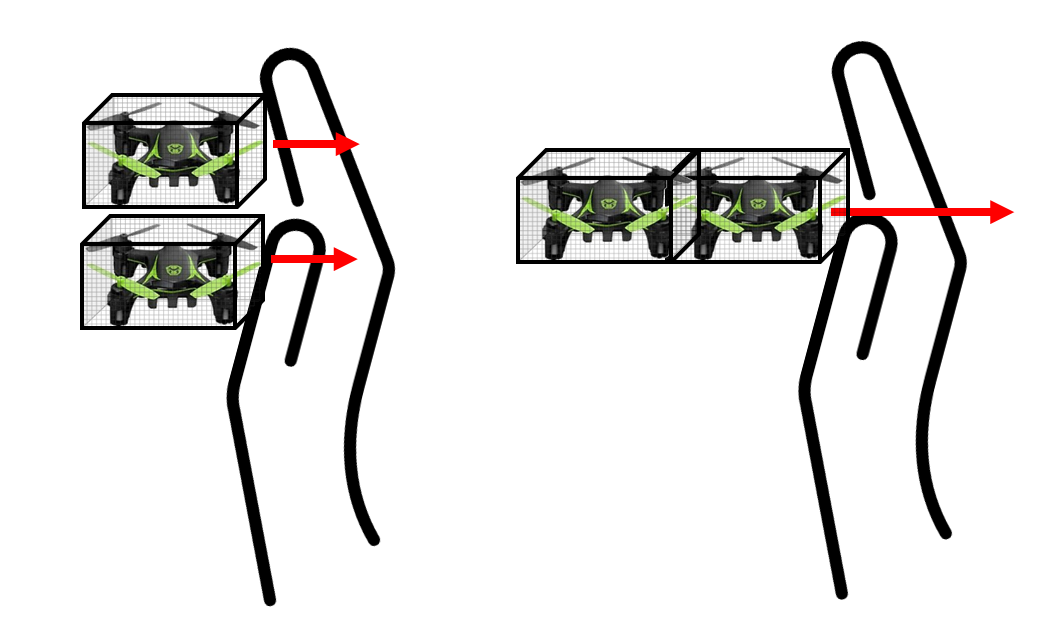}
  \end{center}
  \caption{Force from multiple FLSs. (left) Parallel, (right) Series.}\label{fig:ampforce}
\end{figure}

Therefore, a single FLS will not be capable of providing sufficient force for a salient haptic interaction, especially if we want to generate the kinesthetic sensations necessary for displaying a rigid object. Therefore, the FLSs must coordinate together to combine their thrust so that an amplified force can be applied to the user. This coordination must be done in such a way that the formation remains stable during flight and that force can be passed between drones before eventually being felt by the user. This force amplification can be done either by FLSs in a parallel or series configuration (Fig.~\ref{fig:ampforce}). In parallel, multiple FLSs will contact the user at the same time and will each provide a force proportional to that FLS's disturbance from its own set-point. In series, only a single FLS will be in contact with the user, and both FLSs will have the same disturbance from their set-point; the user feels their combined force.

\noindent{\bf User Safety:}
The goal of our system is to provide non-constrained encountered-type haptic feedback directly to users. 
To do this, the FLS must (1) be modified to allow users to safely have direct contact and (2) react to the user’s gesture input (e.g. push the FLS away from its desired position), and (3) provide meaningful haptic feedback. 
With the first, we use a cage to separate the user's hand from the rotors. Cage shape, weight, and spatial density are key considerations that effect safety, lighting, and flying capability of the FLS. The added weight of the cage would affect the drone’s dynamics and induce imbalance issues when not calibrated. Spatial density of the cage's mesh is important to prevent the user’s finger from accidentally penetrating the surfaces while not affecting the aerodynamics and lighting of the drone too significantly. However, too fine of a spatial density may significantly affect both the lighting and the flying ability of the drone. A balance must be made between choosing a mesh fine enough to protect the user's fingers, but coarse enough to not disrupt the airflow and provide the desired textured lighting.

\noindent{\bf Sensing Users:}
To create an effective touch-based interaction, each FLS should be capable of sensing touch from the user. At a minimum, the FLS should be able to detect the presence or absence of touch, and preferably it will be able to detect the location of touch as well. Capacitive sensors could be integrated into the FLS cage to easily detect the presence of user contact, and a more complex capacitive skin could be developed to determine point of contact. However, determining force of contact from this type of sensor would be challenging, and if this information is required more a more complex mesh of sensors must be developed. Here we propose to use open-loop control for the forces, so direct force sensing from the user is not required. 




%% file: localize.tex
FLSs must localize in order to illuminate a shape and implement haptic interactions.
The modular design of DV will enable an experimenter to evaluate alternative localization techniques.
These techniques may be implemented by hardware components attached to the DV display, an FLS, or both.  
This section presents infrared cameras, RGB cameras, ultra-wideband (UWB) transceivers, retro-reflective markers, or a hybrid of these components for localization.
Below, we describe the four techniques in turn. 

First, one may install 
infrared cameras on the DV display and mount retro-reflective markers on the FLSs.
Software hosted on the DV-Hub will process images captured by the cameras to compute the location of FLSs. 
This centralized technique is inspired by Vicon which provides a high accuracy~\cite{opticalpositioning1,opticalpositioning2,preiss2017}.
It requires the FLSs to maintain a line of sight with the cameras.  
It may be difficult (if not impossible) to form unique markers for more than tens of small drones measuring tens of millimeters diagonally~\cite{preiss2017}.

\begin{figure}
  \begin{center}
    \includegraphics[width=0.9\columnwidth]{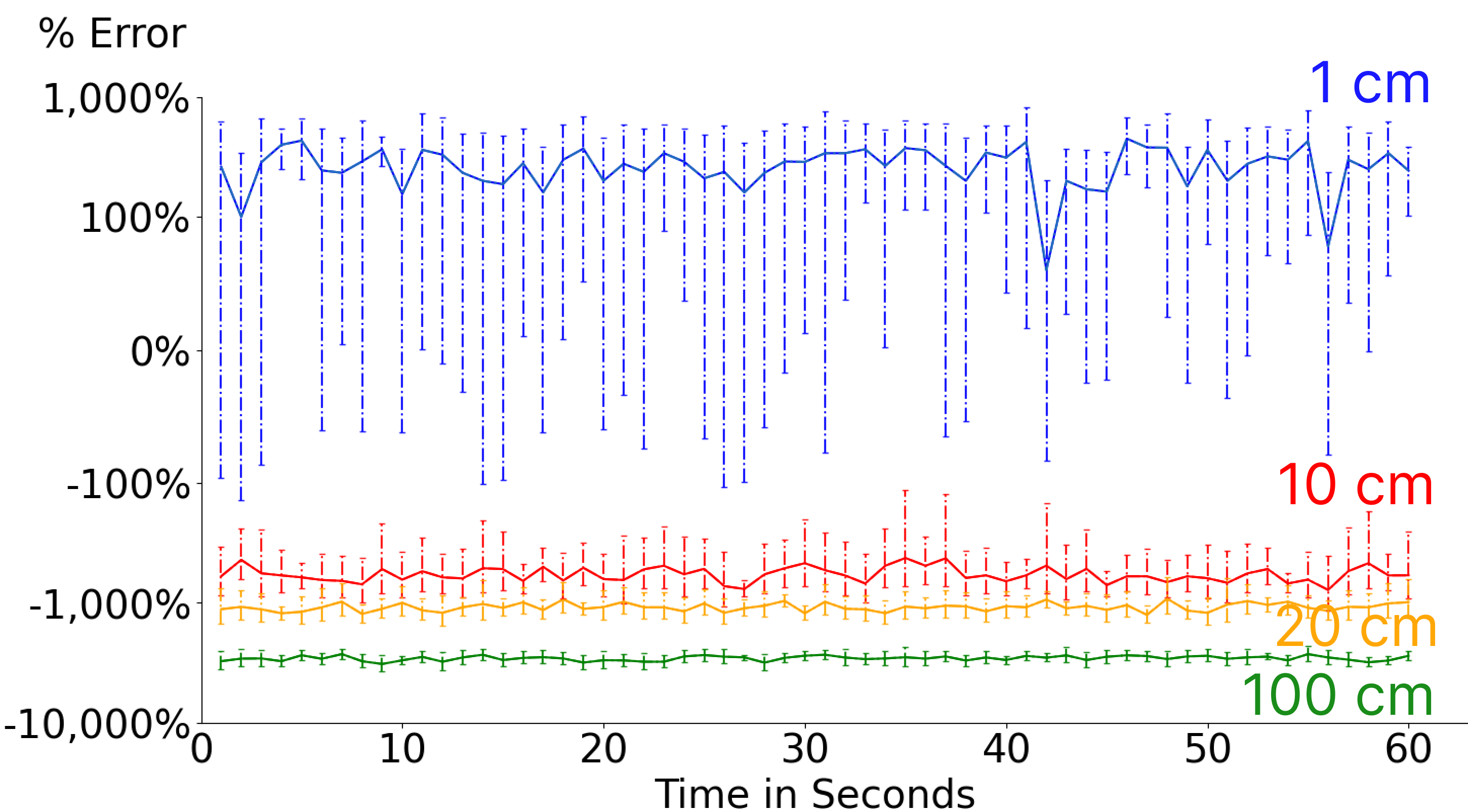}
  \end{center}
  \caption{Percentage error (log scale) in measuring 1 cm distances per second for 60 seconds using two UWB DW1000~\cite{dw1000} cards calibrated at 1, 10, 20, and 100 cm.}\label{fig:UWBerror}
\end{figure}

Second, one may install a set of sensors at known positions~\cite{uwb2015}, e.g., on the side beams, a back panel, floor, or ceiling of the DV display. 
These cards serve as an anchor. 
A sensor may be mounted on each FLS to become a tag. (This is the LOC in Figure~\ref{fig:archfls}.)  Both centralized and decentralized algorithms can be implemented to localize the FLSs using this setup. In this approach, the position of the anchors should be fixed and known by the individual FLSs. This approach requires accurate time synchronization between the anchors.
Depending on the algorithm, time synchronization between anchors and tags may be required as well.
We have experimented with UWB cards~\cite{dw1000} as candidate sensors. 
In our experiments~\cite{imeta2023}, we observed a high margin of error with measuring small distances such as 1 cm, see Figure~\ref{fig:UWBerror}.  
This is consistent with their reported 5-15 cm margin of error in measuring distance~\cite{dw1000,uwb2015,clockdrift22}.

Third, the CPU of each FLS may host a vision-based algorithm that processes the patterns superimposed on the floor (and ceiling) tiles to compute its position~\cite{10.1145/3281548.3281556,4161007,s22155798}.
A limitation of this technique is that some FLSs may occlude the line of sight between other FLSs and the patterns.
With a fix sized DV display, it may be possible to develop algorithms to address this limitation.

Fourth, we envision FLSs with mounted sensors and antennas that enable an FLS to compute its relative distance and angle to an anchor FLS.
A localizing FLS will compare this information with the desired distance and angle in the ground truth, making adjustments to approximate the ground truth more accurately.  
This decentralized  localization algorithm will require line of sight between a localizing FLS and an anchor FLS.
It may implement nature inspired swarming protocols~\cite{reynolds87} to render animated shapes.

%% file: audio.tex
DV audio is a multi-faceted topic.  It includes both how to suppress the unwanted noise from the FLSs and how to generate the acoustics consistent with an application's specifications.
Consider each in turn.
First, the unwanted noise is due to the propulsion system of an FLS consisting of engines and propellers~\cite{Schffer2021}.
This noise increases with higher propeller speeds and larger propeller blades~\cite{Dub2018,sini2013}.

Second, an application may want to render 3D acoustics given the 3D illuminations~\cite{shahram2021}.  For example, consider an illuminated scene where a character placed at one corner of the display shouts at another placed at the farthest corner of the display. The application may want the audio to be louder for those users in close proximity of the shouting character.

We will investigate a host of techniques to address the above two challenges.
With undesirable noise, we will investigate sound suppression and cancellation techniques.
With 3D acoustics, we will investigate use of speakers built in the DV display and a subset of the FLSs.  
We will also consider out-of-the-box solutions. Central to this system would be a pair of noise-canceling headphones, purposefully designed to minimize the ambient drone noise that could potentially distract from the content being displayed. This feature ensures that users are fully focused on the audio-visual presentation, enhancing their overall experience. Additionally, the incorporation of 3D audio technology would allow the sound to dynamically adapt based on the displayed content. This means that as the display changes, the audio component would adjust in real-time, mirroring the spatial positioning and movement of the drones. 

%% file: related.tex
A DV is in the same class of systems as fast 3D printing~\cite{t1000}, Claytronics as physical artifacts using programmable matter consisting of catoms~\cite{matter2005}, Roboxels as cellular robots that dynamically configure themselves into the desired shape and size~\cite{roboxel1993}, 
BitDrones~\cite{gomes2016bitdrones} and GridDrones~\cite{griddrones2018} as interactive nano-drones, and Flying Light Specks (FLSs) as miniature drones with RGB light sources that fly as swarms to illuminate a virtual object~\cite{shahram2021,shahram2022,shahram2022b,mmsys2023}.
These studies use either analytical models, simulation studies, a lab setting, or a hybrid of these to demonstrate their design decisions and to quantify their tradeoffs.
A DV is a much needed experimental testbed to implement and evaluate these techniques for multimedia applications using drones.
Its modular design decisions and haptic user interactions raise many challenging multimedia systems research topics.
If successful, it will contribute to the multimedia content creator's experience in a rich and meaningful fashion while preserving their safety.
It enables them to develop safe applications with complex and intertwined instances of different kinds of information including 3D acoustics, illuminations, and haptic interactions.

The concept of 3D displays using FLSs is introduced in~\cite{shahram2021}.
These FLSs require sensors to measure small distances in order to create 3D formations for visual and haptic displays.
An evaluation of three different types of sensors to measure 1 centimeter distances is presented in~\cite{imeta2023}.
User safety during interaction with a room size display with materialized objects is also important to creating effective interactions; we discuss potential safety considerations in~\cite{shahram2022b}.
An architecture of an FLS system is presented in~\cite{shahram2022}, which introduces key components necessary to realize a successful implementation, including Motill to compute the flight path of FLSs to render a motion illumination, STAG as a battery charging technique, and use of standby FLSs to tolerate FLS failures~\cite{shahram2022}.
This architecture may use the data model of~\cite{mmsys2023} to store the flight paths of an algorithm such as Motill in a file.
A sequence that is displayed repeatedly may read the file instead of executing the resource intensive Motill algorithm~\cite{mmsys2023}.
A DV is a small-scale experimental testbed for FLSs to realize the vision of a Holodeck in science fiction shows and as described in~\cite{shahram2021}.
It builds on these prior studies with its novel use of disaggregated FLSs and a modular framework that emphasizes user safety.
Additionally, a DV uses a swarm of FLSs to amplify force for haptic interactions, creating the important and novel challenges of localization and acoustics.

%% file: dv.tex
A DV is an essential tool to design and implement future multimedia applications using FLSs.  
We are currently designing and developing 
hardware and software 
in support of a DV's kinesthetic haptic user interactions.
This includes IRB-approved human subject studies 
towards the goal of using 
one or more FLSs to provide a high force for generating a stiff surface without losing stability or compromising user safety.

\begin{figure}
  \begin{center}
    \includegraphics[width=0.9\columnwidth]{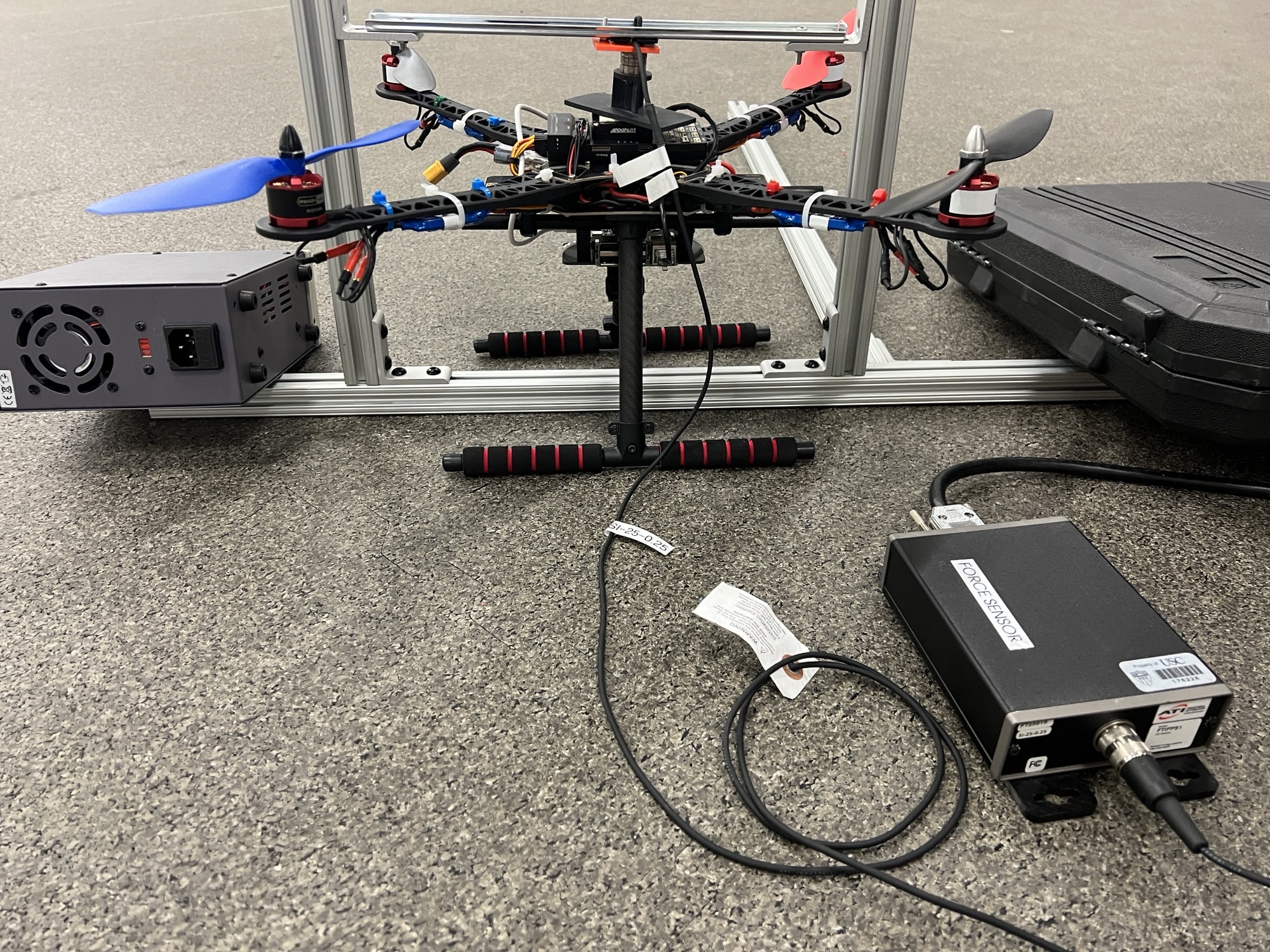}
  \end{center}
  \caption{\href{https://youtu.be/O7nFdFXhbwQ}{Quantifying force} generated by a drone.}\label{fig:measureforce}
\end{figure}

\begin{figure}
\centering
\includegraphics[width=0.9\columnwidth]{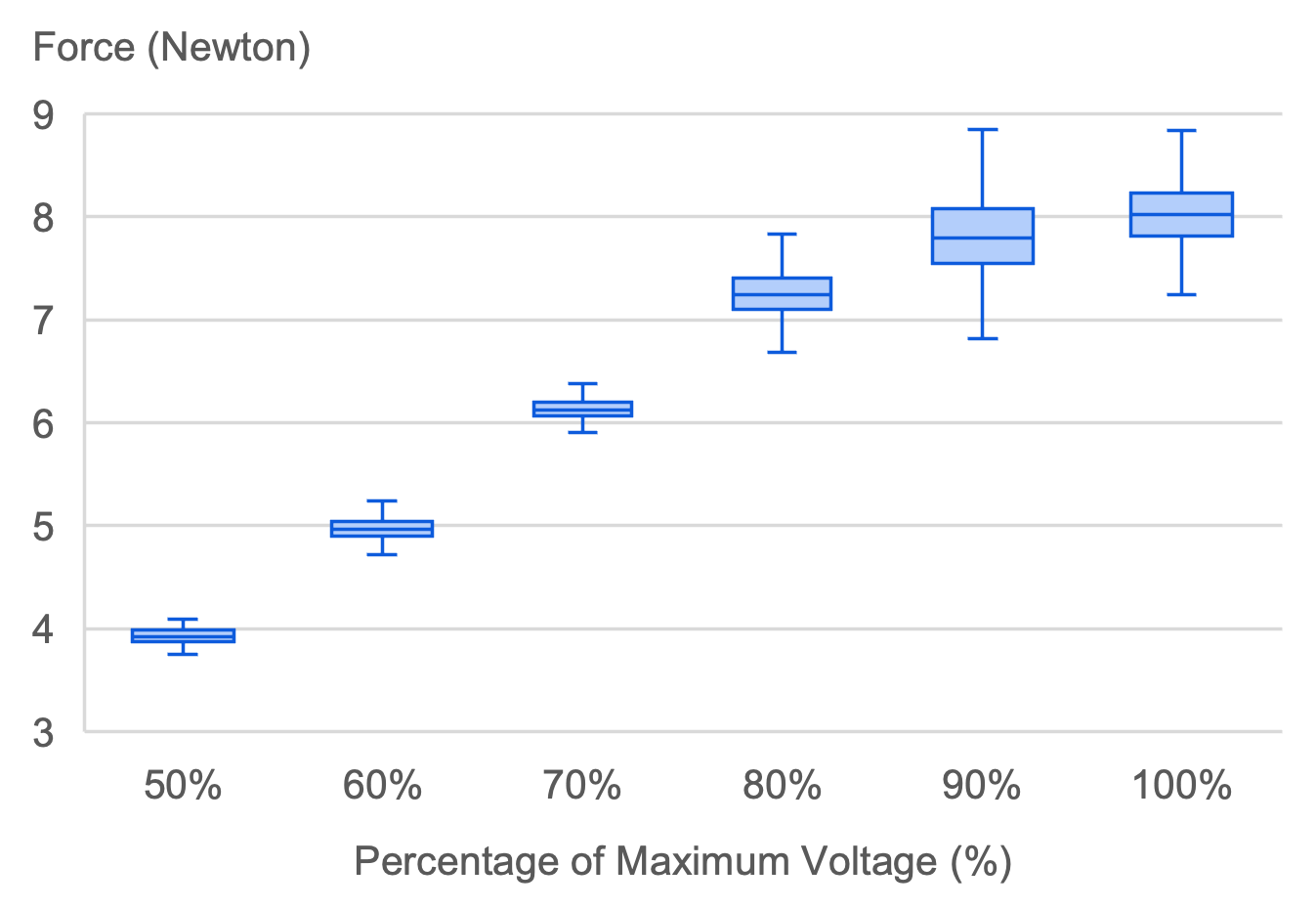}\hfill
\caption{Observed force as a function of the percentage of maximum motor voltage.}
\label{fig:force_thrust}
\end{figure}

Our effort is a combination of empirical studies, physics inspired simulation and analytical models. 
We are designing large drones with different cage arrangements to conduct the human subject studies.
This includes quantifying the  force exerted by the drones, see Figure~\ref{fig:measureforce} and a video demonstration at \url{https://youtu.be/O7nFdFXhbwQ}.
Figure~\ref{fig:force_thrust} shows the measured force in Newtons as a function of the maximum voltage of the motors. The resulting thrust of the motors is directly related to this voltage and creates the measured force.
These measurements were made across 3000 samples during 3 seconds.
The standard deviation is small\footnote{The highest standard deviation is 0.43 and observed with 90\% thrust.}, demonstrating the force exerted can be repeatedly controlled with a high accuracy. 
With our cage designs and a single motor, we observe minimal impact on the thrust and force generated by the motor with the presence of a cage.

Focusing on small drones, 
in addition to the discussions of Section~\ref{sec:interaction}, we are investigating the feasibility of multiple miniature sized FLSs coming in contact with a user at high speeds.
To evaluate the stability of FLSs in such formations, we are conducting empirical studies that fly a swarm of (Crazyflies) drones in a close circular formation.
This formation may be horizontal, vertical, or slanted at 45 degrees.
Videos available at \url{https://youtu.be/oT5RR8RPl0I}, 
\url{https://youtu.be/TQM4hMBwLHM}, and \url{https://youtu.be/NNlWn9VW894} respectively.

We will use the results from both the large and small drones to model how the measured forces change with different sized drones, propeller sizes, and motor characteristics.
Moreover, we are developing a class of PID controllers that allow a drone to follow a pre-defined path and render a pre-specified force output.
These 
use a Vicon localization system 
with centimeter-level accuracy.

We interface a Raspberry Pi 4 as the CPU of Figure~\ref{fig:archfls} with the drone's FC to control its attitude and exerted force.
We are implementing a decentralized localization technique, collision avoidance, and FLS failure handling techniques in software.
Each is a finite state machine with an event-driven framework to implement its functionality.
This framework represents an FLS as an abstract machine that can be in exactly one of a finite
number of states at a given time.
An event handler processes a queue of events that transitions the state of this abstract machine.
It processes events sequentially and atomically, preventing undesirable race conditions caused by inter-leaved execution of events.






Using the Python programming language, we have implemented a scalable emulator with processes. 
Each process represents an FLS, communicates using UDP\footnote{We support both packet loss and out of order packet delivery.}, 
 and implements the aforementioned state machines with an event queue and event handlers.  
At the time of this writing, we have an implementation of a decentralized localization technique, a centralized and decentralized FLS group formation technique, and a centralized (APF~\cite{Sun2020PathPF,saapf2022}) and a decentralized collision avoidance techniques.
While the centralized algorithms are intended for use by the DV Hub, the decentralized algorithms are to be deployed on the FLSs.
The emulator scales both vertically with many (400) cores and horizontally with multiple servers.  
